\documentclass[twocolumn, switch]{article}
\usepackage{preprint}
\usepackage{hyperref}
\usepackage[numbers,square]{natbib}

\usepackage{graphicx}
\usepackage{url}

\hypersetup{colorlinks=true, linkcolor=purple, urlcolor=blue, citecolor=cyan, anchorcolor=black}

\usepackage[utf8]{inputenc}	
\usepackage[T1]{fontenc}
\usepackage{xcolor}
\usepackage{lineno}					
\usepackage{tikz} 					

\usepackage{newfloat}
\DeclareFloatingEnvironment[name={Supplementary Figure}]{suppfigure}
\usepackage{sidecap}
\sidecaptionvpos{figure}{c}

\usepackage{titlesec}
\titlespacing\section{0pt}{12pt plus 3pt minus 3pt}{1pt plus 1pt minus 1pt}
\titlespacing\subsection{0pt}{10pt plus 3pt minus 3pt}{1pt plus 1pt minus 1pt}
\titlespacing\subsubsection{0pt}{8pt plus 3pt minus 3pt}{1pt plus 1pt minus 1pt}

\definecolor{lime}{HTML}{A6CE39}
\DeclareRobustCommand{\orcidicon}{
	\begin{tikzpicture}
	\draw[lime, fill=lime] (0,0)
	circle [radius=0.16]
	node[white] {{\fontfamily{qag}\selectfont \tiny ID}};
	\draw[white, fill=white] (-0.0625,0.095)
	circle [radius=0.007];
	\end{tikzpicture}
	\hspace{-2mm}
}

\title{Jupyter Scatter: Interactive Exploration of Large-Scale Datasets}

\usepackage{authblk}

\author[1]{Fritz Lekschas\href{https://orcid.org/0000-0001-8432-4835}{\orcidicon}}
\author[2]{Trevor Manz\href{https://orcid.org/0000-0001-7694-5164}{\orcidicon}}
\affil[1]{Ozette Technologies, Seattle, WA, USA}
\affil[2]{Harvard Medical School, Boston, MA, US}

\begin{document}

\twocolumn[\begin{@twocolumnfalse}

\maketitle

\begin{abstract}
Jupyter Scatter is a scalable, interactive, and interlinked scatterplot widget
for exploring datasets in Jupyter Notebook/Lab, Colab, and VS Code. Its goal
is to simplify the visual exploration, analysis, and comparison of large-scale
bivariate datasets. Jupyter Scatter can render up to twenty million points,
supports fast point selections, integrates with Pandas DataFrame and
Matplotlib, uses perceptually-effective default settings, and offers a
user-friendly API.\\
\end{abstract}


\vspace{0.5cm}

\end{@twocolumnfalse}]


\section{Summary}

\begin{figure*}[!htbp]
\centering
\includegraphics[width=0.7\linewidth]{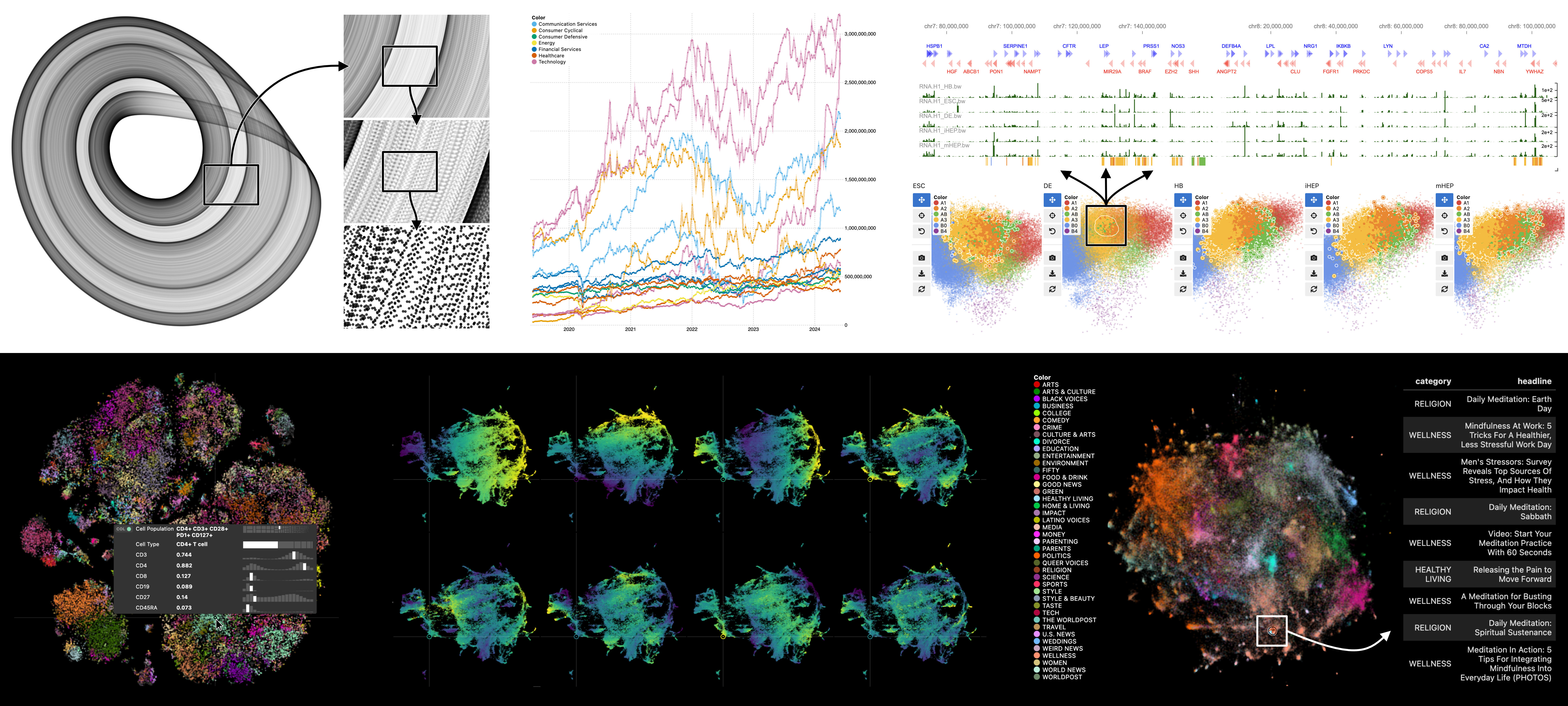}
\caption[]{Examples of Jupyter Scatter. Top row left to right: A 10M point scatterplot of the Roessler Attractor. A connected scatterplot of the market capitalization over the last five years of top ten S\&P500 company according to YCharts. Five linked embedding plots of epigenomic data \citep{dekker2023spatial} that are connected to the HiGlass genome browser \citep{kerpedjiev2018higlass}. Bottom row left to right: A single-cell embedding plot of tumor data \citep{mair2022extricating} that was clustered and annotated with FAUST \citep{greene2022data}. Several linked embedding plots of chromatin state datasets \citep{spracklin2023diverse}. An embedding plot of news headlines \citep{misra2022news} that is linked to a widget for displaying selected articles.}
\label{teaser}
\end{figure*}

Jupyter Scatter is a scalable, interactive, and interlinked scatterplot widget for exploring datasets in Jupyter Notebook/Lab, Colab, and VS Code (Figure~\ref{teaser}). Thanks to its WebGL-based rendering engine \citep{lekschas2023regl}, Jupyter Scatter can render and animate up to several million data points. The widget focuses on data-driven visual encodings and offers perceptually-effective point color and opacity settings by default. For interactive exploration, Jupyter Scatter features two-way zoom and point selections. Furthermore, the widget can compose multiple scatterplots and synchronize their views and selections, which is useful for comparing datasets. Finally, Jupyter Scatter's API integrates with Pandas DataFrames \citep{mckinney2010data} and Matplotlib \citep{hunter2007matplotlib} and offers functional methods that group properties by type to ease accessibility and readability. Extensive documentation and how tos can be found at \href{https://jupyter-scatter.dev}{https://jupyter-scatter.dev} and the code is available at \href{https://github.com/flekschas/jupyter-scatter}{https://github.com/flekschas/jupyter-scatter}.

\section{Usage Scenario}

Jupyter Scatter simplifies the visual exploration, analysis, and comparison of large-scale bivariate datasets. It renders up to twenty million points smoothly, supports fast point selections, integrates with Pandas DataFrame \citep{mckinney2010data}, uses perceptually-effective default encodings, and offers a user-friendly API.

\begin{figure*}[!htbp]
\centering
\includegraphics[width=0.7\linewidth]{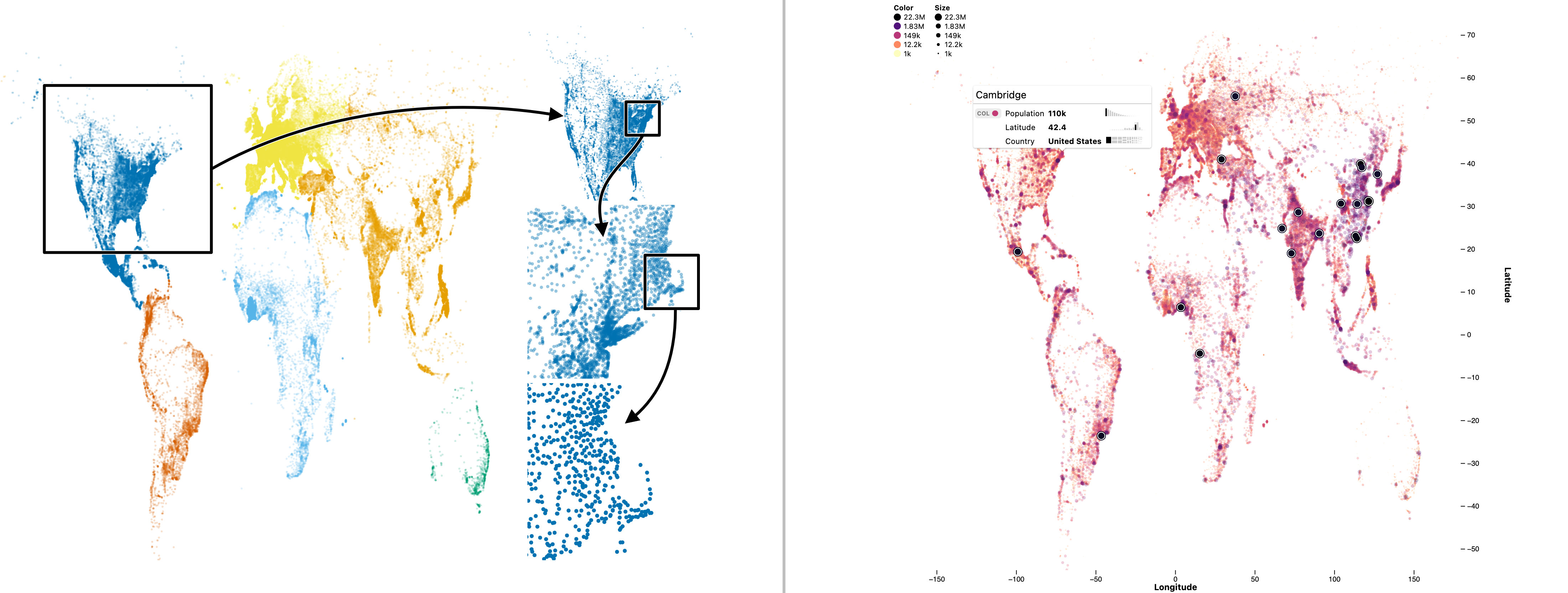}
\caption[]{GeoNames Dataset of Cities Around the World.}
\label{geonames}
\end{figure*}

In the following, we demonstrate its usage for visualizing the GeoNames dataset \citep{geonames}, which contains data about 120k cities world wide. For instance, to visualize cities by their longitude/latitude and color-code them by continent (Figure~\ref{geonames} Left), we create a \texttt{Scatter} widget as follows.

\begin{verbatim}
import jscatter
import pandas as pd

geonames = pd.read_parquet(
  'https://paper.jupyter -scatter.dev/'
    + 'geonames.pq'
)

scatter = jscatter.Scatter(
  data=geonames,
  x='Longitude',
  y='Latitude',
  color_by='Continent',
)
scatter.show()
\end{verbatim}

Without specifying a color map, Jupyter Scatter uses the categorical colorblind-safe palette from \cite{okabe2002color} for the \texttt{Continent} column, which has seven unique values. For columns with continuous data, it automatically selects Matplotlib's \citep{hunter2007matplotlib} \textit{Viridis} color palette. As shown in Figure~\ref{teaser} and Figure~\ref{geonames} Left, Jupyter Scatter dynamically adjusts the point opacity based on the point density within the field of view. This means points become more opaque when zooming into sparse areas and more transparent when zooming out into an area that contains many points. The dynamic opacity addresses over-plotting issues when zoomed out and visibility issues when zoomed in.

Jupyter Scatter offers many ways to customize the point color, size, and opacity encodings. To simplify configuration, it provides topic-specific methods for setting up the scatterplot, rather than requiring all properties to be set during the instantiation of \texttt{Scatter}. For instance, as shown in Figure~\ref{geonames} Right, the point opacity (\texttt{0.5}), size (asinh-normalized), and color (log-normalized population using Matplotlib's \citep{hunter2007matplotlib} \textit{Magma} color palette in reverse order) can be set using the following methods.

\begin{verbatim}
from matplotlib.colors import AsinhNorm, LogNorm
scatter.opacity(0.5)
scatter.size(
  by='Population',
  map=(1, 8, 10),
  norm=AsinhNorm()
)
scatter.color(
  by='Population',
  map='magma',
  norm=LogNorm(),
  order='reverse'
)
\end{verbatim}

To aid interpretation of individual points and point clusters, Jupyter Scatter includes legends, axis labels, and tooltips. These features are activated and customized via their respective methods.

\begin{verbatim}
scatter.legend(True)
scatter.axes(True, labels=True)
scatter.tooltip(
  True,
  properties=['color', 'Latitude', 'Country'],
  preview='Name'
)
\end{verbatim}

The tooltip can show a point's data distribution in context to the whole dataset and include a text, image or audio-based media preview. For instance, the example (Figure~\ref{geonames} Right) shows the distribution of the visually encoded color property as well as the \texttt{Latitude} and \texttt{Country} columns. For numerical properties, the distribution is visualized as a bar chart and for categorical properties the distribution is visualized as a treemap. As the media preview we're showing the city name.

Exploring a scatterplot often involves studying subsets of the points. To select points, one can either long press and lasso-select points interactively in the plot (Figure~\ref{fashion-mnist} Bottom Left) or query-select points (Figure~\ref{geonames} Right) as shown below. In this example, we select all cities with a population greater than ten million.

\begin{verbatim}
scatter.selection(
  geonames.query('Population > 10000000').index
)
\end{verbatim}

The selected cities can be retrieved by calling \texttt{scatter.selection()} without any arguments. It returns the data record indices, which can then be used to get back the underlying data records.

\begin{verbatim}
cities.iloc[scatter.selection()]
\end{verbatim}

To automatically register changes to the point selection one can observe the \texttt{scatter.widget.selection} traitlet. The observability of the selection traitlet (and many other properties of \texttt{scatter.widget}) makes it easy to integrate Jupyter Scatter with other Jupyter Widgets.

For instance, Figure~\ref{fashion-mnist} (Left) shows a UMAP \citep{leland2018umap} embedding of the Fasion MNIST dataset \citep{xiao2017fashion} where points represent images and the point selection is linked to an image widget that loads the selected images.

\begin{verbatim}
import ipywidgets
import jscatter

fashion_mnist = pd.read_parquet(
    'https://paper.jupyter -scatter.dev/'
      + 'fashion -mnist -embeddings.pq'
)

# Custom image widget
images = ImagesWidget()

scatter = jscatter.Scatter(
  data=fashion_mnist,
  x='umapX', y='umapY',
  color_by='class',
  background_color='black',
  axes=False,
)

ipywidgets.link(
  (scatter.widget, 'selection'),
  (images, 'images')
)

ipywidgets.AppLayout(
  center=scatter.show(),
  right_sidebar=images
)
\end{verbatim}

Comparing two or more related scatterplots can be useful in various scenarios. For example, with high-dimensional data, it might be necessary to compare different properties of the same data points. Another scenario involves embedding the high-dimensional dataset and comparing different embedding methods. For large-scale datasets, it might be useful to compare different subsets of the same dataset or entirely different datasets. Jupyter Scatter supports these comparisons with synchronized hover, view, and point selections via its \texttt{compose} method.

\begin{figure*}[!htbp]
\centering
\includegraphics[width=0.7\linewidth]{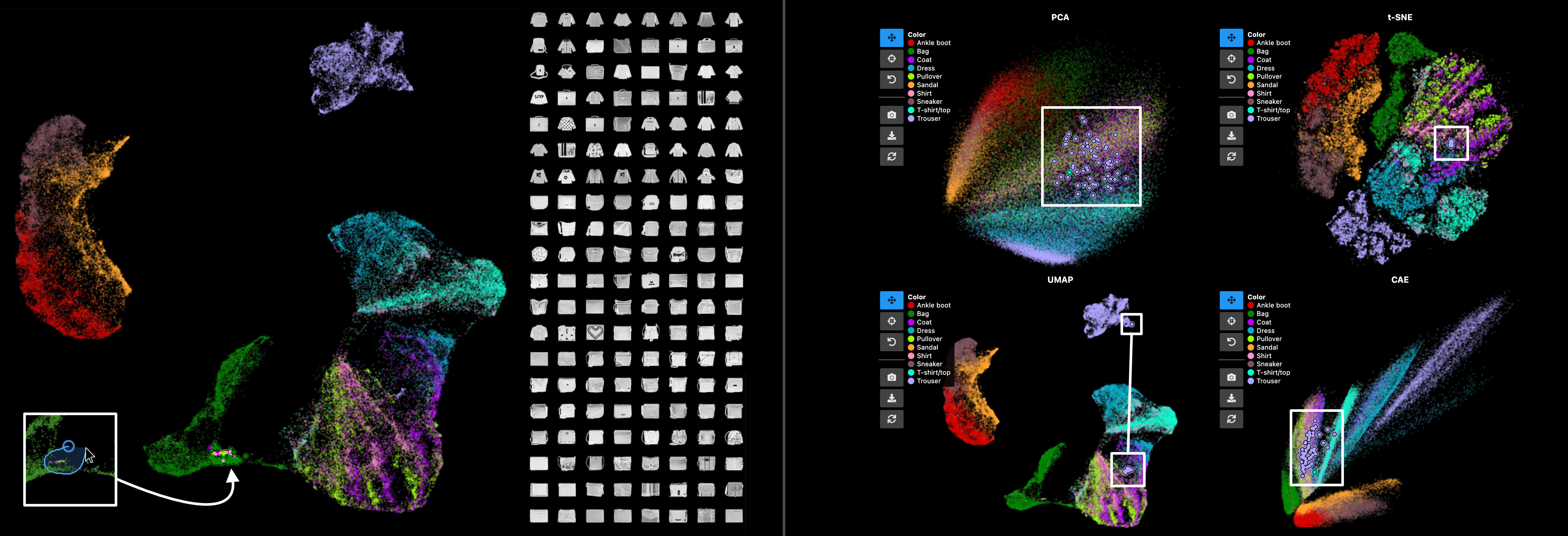}
\caption[]{Fashion MNIST Embeddings. Left: Integration of Jupyter Scatter with an image widget through synchronized point selections. Right: Four scatterplots with synchronized point selection.}
\label{fashion-mnist}
\end{figure*}

For instance, there are many ways to embed points into two dimensions, including linear and non-linear methods, and comparing point clusters between different embedding methods can be insightful. In the following, we compose a two-by-two grid of four embeddings of the Fashion MNIST dataset \citep{xiao2017fashion} created with PCA \citep{pearson1901}, UMAP \citep{leland2018umap}, t-SNE \citep{vandermaaten2008visualizing}, and a convolutional autoencoder \citep{kingma2013auto}. As illustrated in Figure~\ref{fashion-mnist} (Right), the point selection of the four scatterplots is synchronized.

\begin{verbatim}
from jscatter import Scatter, compose
config = dict(
  data=fashion_mnist,
  color_by='class',
  legend=True,
  axes=False,
  zoom_on_selection=True,
)

pca = Scatter(x='pcaX', y='pcaY', **config)
tsne = Scatter(x='tsneX', y='tsneY', **config)
umap = Scatter(x='umapX', y='umapY', **config)
cae = Scatter(x='caeX', y='caeY', **config)

compose(
  [
    (pca, "PCA"),
    (tsne, "t -SNE"),
    (umap, "UMAP"),
    (cae, "CAE")
  ],
  sync_selection=True,
  sync_hover=True,
  rows=2,
)
\end{verbatim}

Note, by setting \texttt{zoom\_on\_selection} to \texttt{True} and synchronizing selections, selecting points in one scatter will automatically select and zoom in on those points in all scatters.

\section{Statement of Need}

Jupyter Scatter is primarily a tool for data scientists to visually explore and compare bivariate datasets. Its ability for two-way point selections and synchronized plots, enable interactive exploration and comparison in ways that is not possible with existing widgets (e.g., multiple linked scatterplots) or requires considerable effort to set up (e.g., two-way communication of point selections).

Further, due to its usage of traitlets \citep{traitlets}, Jupyter Scatter integrates easily with other widgets, which enables visualization researchers and practitioners to build domain-specific applications on top of Jupyter Scatter. For instance, the \textit{Comparative Embedding Visualization} widget \citep{manz2024general} uses Jupyter Scatter to display four synchronized scatterplots for guided comparison of embedding visualizations. \href{https://co-labo.org/}{Andrés Colubri's research group} is actively working on a new version of their \textit{Single Cell Interactive Viewer} which will be based on Jupyter Scatter.

\section{Implementation}

Jupyter Scatter has two main components: a Python program running in the
Jupyter kernel and a front-end program for interactive visualization. The
Python program includes a widget and an API layer. The widget defines the view
model for drawing scatterplots, while the API layer simplifies defining the
view model state, integrating with Pandas DataFrames \citep{mckinney2010data} and
Matplotlib \citep{hunter2007matplotlib}. The front-end program is built on top of
regl-scatterplot \citep{lekschas2023regl}, a high-performance rendering library
based on WebGL, ensuring efficient GPU-accelerated rendering.

All components are integrated using anywidget \citep{anywidget} to create a
cross-platform Jupyter widget compatible with various environments, including
Jupyter, JupyterLab, Google Colab, VS Code, and dashboarding frameworks like
Shiny for Python, Solara, and Panel. The Python program uses anywidget and
ipywidgets \citep{ipywidgets} to commuincate with the front-end, using binary data
support to efficiently send in-memory data to the GPU, avoiding the overhead of
JSON serialization. This approach enables the transfer of millions of data
points from the Python kernel to the front-end with minimal latency.
Bidirectional communication ensures the visualization state is shared between
the front-end and kernel, allowing updates to scatterplot properties and access
to states like selections. Coordination is managed using anywidget APIs,
enabling connections to other ipywidgets like sliders, dropdowns, and buttons
for custom interactive data exploration widgets.

\section{Related Work}

There are many Python packages for rendering scatterplots in notebook-like environments. General-purpose visualization libraries like Matplotlib \citep{hunter2007matplotlib}, Bokeh \citep{bokeh}, or Altair \citep{vanderplas2018altair} offer great customizability but do not scale to millions of points. They also don't offer bespoke features for exploring scatterplots and require manual configuration.

More bespoke dataset-centric plotting libraries like Seaborn \citep{waskom2021seaborn} or pyobsplot \citep{pyobsplot} require less configuration and make it easier to create visually-pleasing scatterplots but they still fall short in terms of scalability.

Plotly combines great customizability with interactivity and can render scatterplots of up to a million points. However, drawing many more points is challenging and the library also focuses more on generality than dedicated features for scatterplot exploration and comparison. Plotly's WebGL rendering mode is also bound to the number of WebGL contexts your browser supports (typically between 8 to 16) meaning that it can't reader more 8 to 16 plots when using the WebGL render mode. Jupyter Scatter does not have this limitation as it uses a single WebGL renderer for all instantiated widgets, which is sufficient as static figures don't need constant re-rendering and one will ever only interact with a single or few plots at a time. Being able to render more than 8 to 16 plots can be essential in notebook environments as these are often used for exploratory data analysis.

Datashader \citep{datashader} specializes on static rendering of large-scale datasets and offers unparalleled scalability that greatly exceeds that of Jupyter Scatter. One can also fine-tune how data is aggregated and rasterized. However, this comes at the cost of limited interactivity. While it's possible to interactively zoom into a rasterized image produced by Datashader, the image is just drawn at scale instead of being re-rendered at different field of views. Re-rendering can be important though to better identify patters in subsets of large scatterplots through optimized point size and opacity.

Finally, except for Plotly, none of the tools support interactive point selections, a key feature of Jupyter Scatter. Also, no other library offers direct support for synchronized exploration of multiple scatterplots for comparison.

\section*{Acknowledgements}
\footnotesize
We acknowledge and appreciate contributions from Pablo Garcia-Nieto,
Sehi L'Yi, Kurt McKee, and Dan Rosén. We also thank Nezar Abdennur for his
feedback on the initial API design.
\normalsize

\bibliography{main.bib}

\end{document}